# Spin-polarized imaging of the antiferromagnetic structure and field-tunable bound states in kagome magnet FeSn

Hong Li[1], He Zhao[1], Qiangwei Yin[2], Qi Wang[2], Zheng Ren[1], Shrinkhala Sharma[1], Hechang Lei[2], Ziqiang Wang[1], Ilija Zeljkovic[1,*]

[1] Department of Physics, Boston College, Chestnut Hill, MA 02467

[2] Department of Physics and Beijing Key Laboratory of Opto-electronic Functional Materials & Micro-nano Devices, Renmin University of China, Beijing 100872, China

*corresponding author (ilija.zeljkovic@bc.edu)

## Abstract

**Kagome metals are as an exciting playground for the explorations of novel phenomena at the intersection of topology, electron correlations and magnetism. The family of FeSn-based kagome magnets in particular attracted a lot of attention for simplicity of the layered crystal structure and tunable topological electronic band structure. Despite a significant progress in understanding their bulk properties, surface electronic and magnetic structures are yet to be fully explored in many of these systems. In this work, we focus on a prototypical kagome metal FeSn. Using a combination of spin-averaged and spin-polarized scanning tunneling microscopy, we provide the first atomic-scale visualization of the layered antiferromagnetic structure at the surface of FeSn. In contrast to the field-tunable electronic structure of cousin material $Fe_3Sn_2$ that is a ferromagnet, we find that electronic density-of-states of FeSn is robust to the application of external magnetic field. Interestingly, despite the field insensitive electronic band structure, FeSn exhibits bounds states tied to specific impurities with large effective moments that strongly couple to the magnetic field. Our experiments provide microscopic insights necessary for theoretical modeling of FeSn and serve as a spring board for spin-polarized measurements of topological magnets in general.**

## Introduction

Quantum materials composed of atoms arranged on a lattice of corner-sharing triangles (kagome lattice) are a versatile platform to explore electronic phenomena at the intersection of band topology and electronic correlations [1–9]. While the initial excitement behind these systems stemmed from the possibility of realizing spin liquid phases [1,10], recent experiments revealed a range of other novel electronic phases that can emerge on a kagome lattice in the presence of spin-orbit coupling, non-trivial Berry curvature and/or magnetism. These for example include topological flat bands [11,12], Chern magnet phase [13], Weyl semimetal phase and Fermi arcs [14,15], and various density waves [16–21].

In the pursuit of exotic electronic phenomena, the family of $Fe_xSn_y$ kagome magnets has been of particular interest [22–31]. Materials in this family are characterized by the prototypical electronic band structure associated with the kagome lattice, consisting of Dirac cones at the Brillouin zone boundary and a dispersionless flat band [24,25,27,28,30]. These systems exhibit a layered crystal structure composed of different sequences of $Fe_3Sn$ kagome layers and honeycomb Sn layers stacked along the *c*-axis. This stacking order directly influences the type of emergent magnetic ordering in the bulk [22,32,33]. For example, $Fe_3Sn_2$, composed of $Fe_3Sn$-$Fe_3Sn$-Sn building blocks is ferromagnetic [24,25,30,31]. On the other hand FeSn, composed of alternating $Fe_3Sn$ layers and Sn layers is a layered antiferromagnet: Fe spins within each layer align ferromagnetically, but couple antiferromagnetically between adjacent layers [34] (Fig. 1a,h). Despite the well-known magnetic structures in the bulk, magnetic ordering at the surface of Fe-based kagome metals and its tunability with external perturbations is yet to be fully investigated. Experimentally establishing this would be essential for several reasons. First, given the broken crystal symmetry at the surface, the magnetic structure may be different than that in the bulk. Dichotomy between surface and bulk magnetism has indeed been hypothesized to occur in other magnetic topological systems [35]. Second, surface magnetization can lead to the transition from massless to massive Dirac fermions [25], the latter of which in principle carrying a non-trivial Chern number. As such, direct measurement of magnetic properties at the surface is highly desirable for a complete understanding of these materials. However, such measurements have been challenging to achieve in many of the kagome magnets to-date. In this work, we use spin-polarized scanning tunneling microscopy and spectroscopy to visualize the layered antiferromagnetic structure at the surface of prototypical kagome metal FeSn.

**Results**

Kagome metal FeSn is a bulk antiferromagnet (Neel temperature $T_N \approx$ 370 K [32]) characterized by the P6/mmm group symmetry and a hexagonal lattice ($a$ = 5.298 Å and $c$ = 4.448 Å) [36]. Its crystal structure consists of alternating honeycomb Sn layers and $Fe_3Sn$ kagome layers (Fig. 1a,b). We cleave bulk single crystals of FeSn in ultra-high vacuum (UHV) and immediately insert them into the STM head where they are imaged at 4.5 K (Methods). STM topographs reveal both possible surface terminations: honeycomb Sn and kagome $Fe_3Sn$ (Fig. 1c,d). While both of them exhibit a similar hexagonal structure with an in-plane lattice constant of $a \approx$ 5.3 Å (Fig. 1c,d), they are characterized by distinct spectroscopic signatures: one with a spectral peak near -50 meV (Fig. 1e) and the other where the peak is absent (Fig. 1f). We identify each termination based on the following. First in Fig. 1c, we can clearly discern atoms arranged on a honeycomb lattice, which is consistent with individual atoms in the complete Sn layer (inset in Fig. 1c) and qualitatively similar to the STM topographs of single layer stanene [37]. Second, partial overlayers of Sn on top of the kagome layer, such as the one seen in Fig. 1g,i, have also been reported to occur on isostructural CoSn [38], thus pointing again towards the taller termination indeed being the Sn layer.

We first explore the Sn surface termination, which we have predominantly observed in our measurements. We focus on a region encompassing two Sn terraces across a single unit cell step (Fig. 2a,b). Using a conventional (spin-averaged) STM tip, we find that both terraces show identical dI/dV spectra (Fig. 2c,d). To evaluate if the electronic band structure changes with applied magnetic field B, as it often does in magnetic materials [25,39], we repeat the dI/dV measurement as a function of out-of-plane magnetic field. We find no difference between zero-field dI/dV spectra and those acquired in +/-4 T magnetic field (minus sign denotes the reversal of the magnetic field applied antiparallel to the c-axis) (Fig. 2e). To rule out the unlikely scenario that the cleaving process affects the FeSn surface properties, we demonstrate the same absence of magnetic field tunability of dI/dV spectra in home-grown FeSn thin films synthesized by molecular beam epitaxy (Supplementary Figure 1).

Magnetic ordering in FeSn lifts the degeneracy of the electronic bands, leading to spin majority and minority bands that can be observed on both surface terminations [40]. Due to the antiferromagnetic coupling between adjacent layers stacked along the *c*-axis (Fig. 1h), spin majority and minority bands should in principle "switch" between adjacent layers. However, as shown in Figure 2, conventional STM cannot resolve the difference in the density of states between inequivalent terraces. To explore the spin texture in more detail, we use spin-polarized STM (Methods), a valuable tool for real-space spin-resolved imaging of various antiferromagnets, such as Fe-based metals including superconductors [41–44], Ir-based oxides [45,46] and elemental Cr [47,48] and Mn [49]. We again locate a region with several consecutive steps, each between adjacent Sn layers, and acquire a dI/dV linecut across the steps (Fig. 3a,b). For the ease of discussion, we label each terrace with consecutive integers, starting with 1 on the lowest terrace. We find that the average dI/dV spectra acquired on terraces denoted by even numbers are all identical, but markedly different from the spectra acquired on odd terraces (Fig. 3c). In particular, the spectral peak at negative energy exhibits a pronounced spectral weight shift between the two types of terraces. This trend is present across all steps imaged, and can be visualized as the systematic variation in differential conductance at negative energies (Fig. 3d,g). Since dI/dV spectra acquired using a spin-averaged tip show no difference between consecutive steps, the difference observed here can be understood as a consequence of spin-polarized tunneling. In principle, for a fixed tip-sample distance, the tunneling current will depend on the overlap between the spin direction of the tip and that of the sample [50]. In our experiment, as the spin orientation of the tip remains the same across all terraces, the sample spin direction has to be different between adjacent terraces (Fig. 3f), which is reflected in the measured dI/dV spectrum. Systematic variation of dI/dV spectra is consistent with the expected layered antiferromagnetic structure, where neighboring terraces offset by a full unit cell step height should have spins polarized in opposite directions in the *ab*-plane (Fig. 1h). This interpretation is further confirmed by the magnetic field dependence, where the difference between the two types of terraces is reversed as the spin of the tip is "flipped" by external field (Fig. 3e). Our SP-STM measurements establish the existence of layered antiferromagnetism at the surface of FeSn, consistent with the bulk antiferromagnetic order.

To demonstrate that this behavior is not confined to a single surface termination, we show dI/dV spectra acquired on top of the occasionally observed $Fe_3Sn$ kagome surface terraces (Fig. 4). dI/dV spectra acquired using a spin-polarized STM tip across the consecutive $Fe_3Sn$ steps show pronounced spectral differences between them across the entire energy range imaged (Fig. 4d). This is in contrast to the Sn termination, where the difference in spin-polarized integrated density of states primarily occurs on the negative side, but dI/dV spectra at positive energies appear nearly indistinguishable (Fig. 4c).

It is interesting to note that the surface of cleaved single crystals of FeSn also shows several types of impurities with distinct spatial signatures (Fig. 5a): two-fold symmetric (labeled A, B and C), $C_3$-symmetric (labeled D) or $C_6$-symmetric (labeled E). Based on the intra-unit cell position with respect to the Sn surface atoms, we can identify impurities A, B and C to occur at the Fe site, while D and E occur at the Sn site (Fig. 5b). The Fe-site location of A, B and C defects would naturally explain the $C_2$-symmetric signature due to the two-fold symmetry of the atomic arrangement around this particular Fe site that is shared between two neighboring triangles (Fig. 5b). We note that this two-fold electronic signature is not indicative of a nematic phase, seen in cousin kagome system $Fe_3Sn_2$ [25], since the spatial signature rotates by multiple of 120 degrees for different impurities (see different orientation of A, B and C in Fig. 5a), but simply a consequence of the crystal structure. In addition to the two-fold electronic signal, impurity A shows bound states located in close proximity to zero energy (Fig. 5c,d,g). These can be modulated by the application of external magnetic field across the Fermi level for a moderate range of field values used in our experiment (Fig. 5e,f). Interestingly, the bound states evolve in different directions as the direction of the magnetic field is reversed, indicating a fixed local moment irrespective of the field direction (Fig. 5e,f). Field-tunable impurity states have previously been reported at defect sites in kagome magnet $Co_3Sn_2S_2$ [51,52]. In contrast to the bound states in $Co_3Sn_2S_2$ however, field-tunable bound states observed in FeSn are in close proximity to the Fermi level.

## Conclusion and discussion

Our SP-STM experiments reveal staggered modulations of dI/dV spectra across consecutive surface terraces, consistent with a robust layered antiferromagnetic structure of FeSn that persists at the surface. We note that it is difficult to conclusively identify the spectral features in dI/dV spectra in relation to particular electronic bands given the complexity of the band structure [27], and this is beyond the scope of the current paper. Given that theoretical calculations indicate the presence of several van Hove singularities in the vicinity of the Fermi level [27], it is conceivable that the peak in dI/dV around -50 meV in Fig. 2 can be attributed to a van Hove singularity, but further work will be necessary to elucidate this. We deem that the origin of this spectral peak is unlikely to arise due to a flat band, as the flat band is located more than 200 meV below the Fermi level [27]. We can also rule out impurity bound states as dI/dV spectra are spatially extremely homogeneous (Fig. 2c). We further mention that although the kagome plane is located below the Sn layer, termination-dependent band structure calculations and ARPES measurements [27] indicate that some Fe bands should still be detectable on the Sn termination. We find that electronic band structure is largely insensitive to the application of moderate out-of-plane magnetic field, in contrast to the ferromagnetic cousin $Fe_3Sn_2$ where the electronic band structure rapidly evolves with field as the magnetic moments rotate out-of-plane [25]. Field insensitivity of the band structure observed here is also different from another kagome antiferromagnet YMn$_6$Sn$_6$, where magnetic field leads to a continuous filed-induced band renormalization [53] attributed to a combination of spin canting and orbital magnetization. This could suggest that these effects are negligible in FeSn. Despite insensitivity of the band structure to magnetic field, we reveal that certain Fe-site impurities harbor bound states tunable by magnetic field, shifting to higher energy regardless of the direction of magnetic field. The demonstrated ability to shift the energy of these bounds states away and across the Fermi level could potentially be harnessed in transport measurements if a sufficient density of impurities is induced during the growth process.

## Methods

Bulk single crystals of FeSn are grown using the self-flux method. The high-purity Fe (piece) and Sn (shot) were put into corundum crucibles and sealed into quartz tubes with a ratio of Fe:Sn = 2:98. The tube was heated to 1273 K and held there for 12 h, then cooled to 823 K at a rate of 3 K/h. The flux was removed by centrifugation, and shiny crystals with typical size about 2 x 2 x 5 mm$^3$ can be obtained.

FeSn single crystals are cleaved in UHV in about 10$^{-10}$ Torr pressure and immediately inserted into the STM head. For spin-averaged measurements, we use a chemically-etched W tip, annealed in vacuum to remove the oxide layer from the surface.  To create spin-polarized tips, we start with the same etched W wire, but train it on the FeSn sample surface by fast scanning and bias pulsing. The tip can ultimately become spin-polarized, likely by picking up one or more magnetic atoms from the sample surface. In order to demonstrate that the tip is actually spin-polarized, we test the tip on the cleaved surface of FeTe single crystal after the completion of measurements on FeSn (Supplementary Figure 2). If the tip is spin-polarized, it will show stripe-like signature of antiferromagnetic ordering in FeTe, as for example reported in Refs. [41]. As shown in Fig. S2, our spin-polarized tip shows the desired stripe modulation related to the underlying antiferromagnetism in FeTe. The tips we create in this manner are "ferromagnetic": the polarization can be flipped by external magnetic field.  We test this by the dependence of STM topographs in external magnetic field applied perpendicular to the sample surface. For example, STM topographs acquired with such a tip show almost no difference in 0 T and -1 T magnetic field (tip has same polarization in both fields), but exhibit a stripe "shift" between 1T and 0T (tip polarization direction is "flipped"), demonstrating that the W tip is indeed spin-polarized (Supplementary Figure 2).

## Data availability

The datasets generated and/or analyzed during the current study are available in the Zenodo repository, https://doi.org/10.5281/zenodo.6456564. All other data that support the findings of this study are available from the corresponding author upon reasonable request.

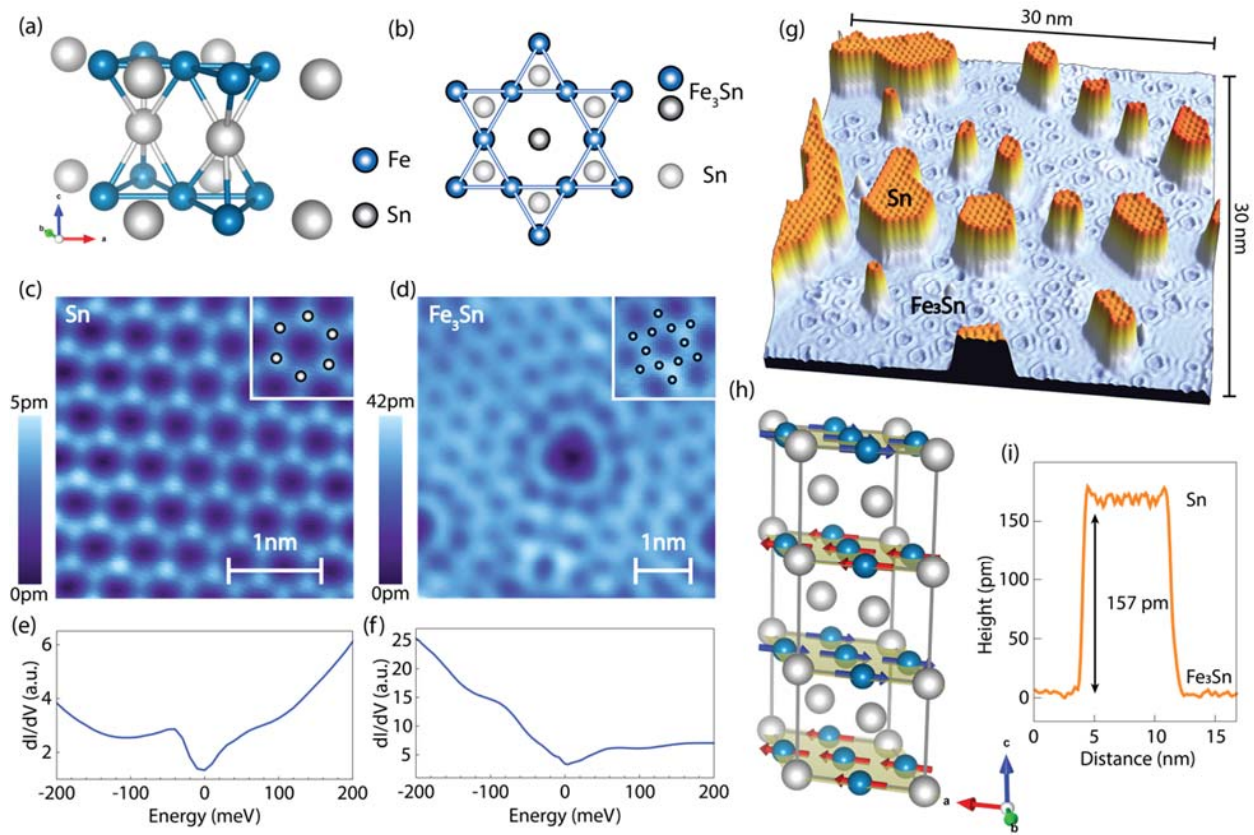

**Figure 1. FeSn crystal structure and different surface terminations. (a)** A 3D schematic of the FeSn atomic structure. **(b)** The *ab*-plane schematic of the two possible surface terminations: the kagome $Fe_3Sn$ layer and the hexagonal Sn layer. Blue (gray) spheres in (a,b) represent Fe (Sn) atoms. **(c, d)** Zoomed-in STM topographs of (c) the Sn surface termination and (d) the $Fe_3Sn$ surface termination. Insets in (c,d) show the arrangement of surface atoms on top of the topograph. **(e, f)** Average dI/dV spectra acquired on (e) the Sn surface and (f) the $Fe_3Sn$ surface shown in (c, d). **(g)** An STM topograph showing Sn islands (upper layer) on top of the $Fe_3Sn$ surface (lower layer). **(h)** A 3D model showing the layered antiferromagnetic structure of FeSn. Red and blue arrows represent spins that are polarized in the *ab*-plane in opposite directions. **(i)** A linecut across a small Sn island, showing the apparent Sn-$Fe_3Sn$ step height of 157 pm, which is a bit smaller compared to the expected height of 222 pm. STM setup condition: (c) $I_{set}$ = 600 pA, $V_{sample}$ = 100 mV; (d) $I_{set}$ = 800 pA, $V_{sample}$ = -60 mV; (e) $I_{set}$ = 800 pA, $V_{sample}$ = 400 mV, $V_{exc}$ = 5 mV; (f) $I_{set}$ = 800 pA, $V_{sample}$ = 200 mV, $V_{exc}$ = 2 mV.

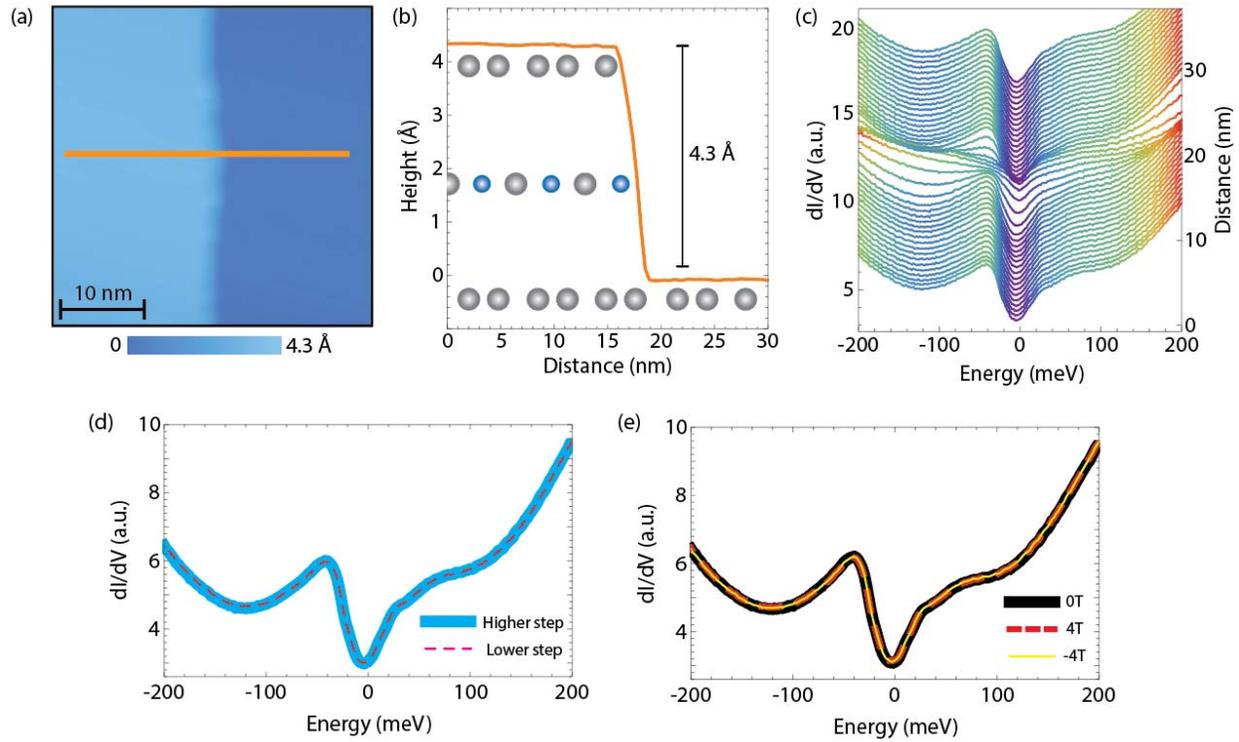

**Fig 2. Absence of electronic band structure tunability with magnetic field. (a)** STM topograph across a step between two Sn terraces. **(b)** Schematic of the step in (a) with a topographic profile along orange line in (a). **(c)** Waterfall plot of dI/dV spectra along the orange line in (a), showing uniformity of spectra away from the edge that appear indistinguishable on either terrace. **(d)** Average dI/dV spectra on the two terraces in (a), overlapping one another almost exactly. **(e)** Average dI/dV spectra acquired on the higher terraces in (a) under 0 T and ± 4 T magnetic field applied perpendicular to the sample surface. All three spectra again appear indistinguishable. STM setup condition: (a) $I_{set}$ = 10 pA, $V_{sample}$ = 1V; (c-e) $I_{set}$ = 800 pA, $V_{sample}$ = 200 mV, $V_{exc}$ = 2 mV.

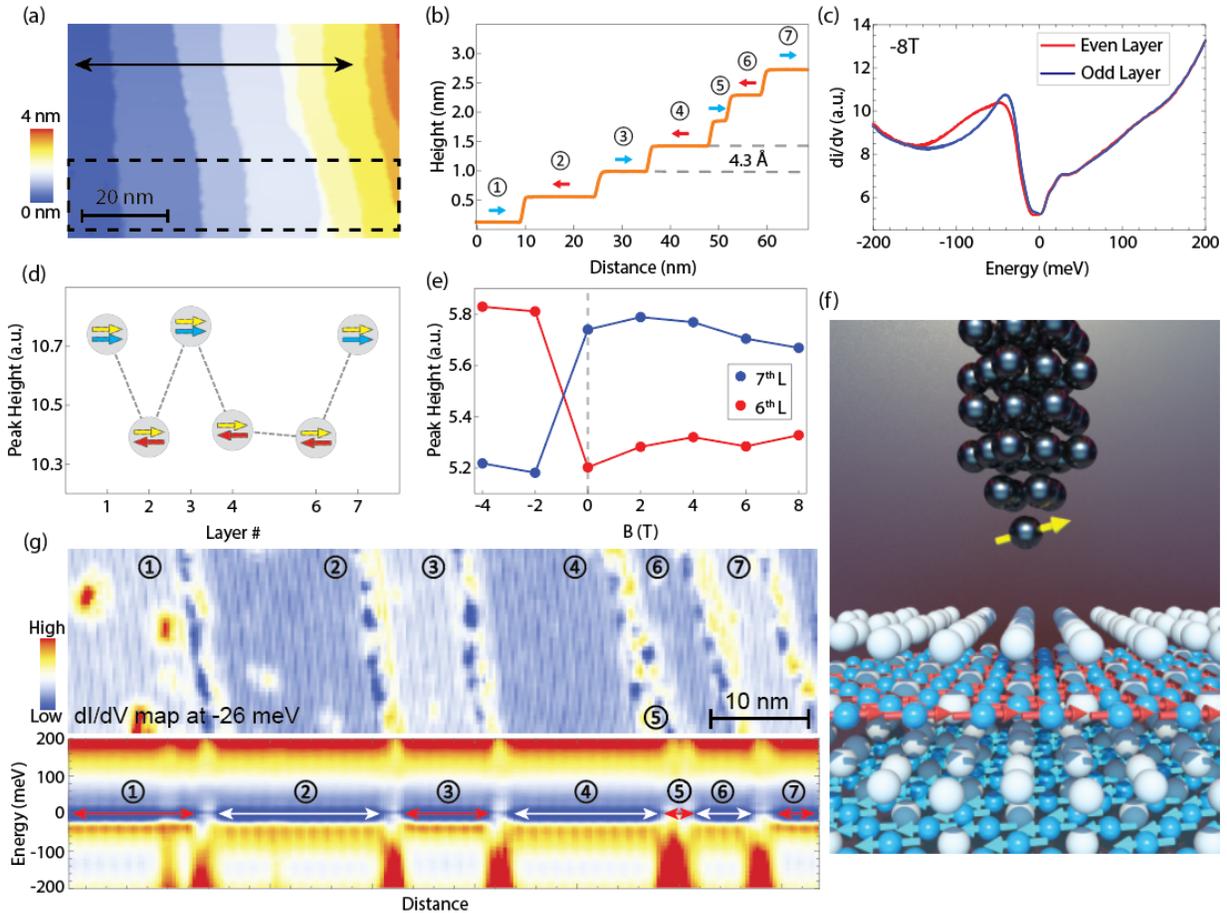

**Figure 3. Spin-polarized imaging the layered antiferromagnetic structure of the Sn layer. (a)** STM topograph of a region containing 7 unit cell step heights using a spin-polarized STM tip. **(b)** A topographic line profile along the black line in (a). The red and blue arrows in (b) denote the spins in each step. **(c)** Average dI/dV spectra taken over even (red curve) and odd (blue curves) steps in (a) under -8 T magnetic field applied perpendicular to the surface. Note that 1st, 3rd, 7th layer have exactly the same spectra (the blue one); 2nd, 4th, 6th layers also share same spectrum (red one). **(d)** The spectral peak height at ~ -50 meV as a function of the layer number marked in (b). Yellow arrow stands for tip spin polarization, while the blue and red arrows stand for odd layer and even layer spin polarizations respectively. **(e)** Spectral peak height plotted as a function of applied magnetic field magnitude for the 6th layer (red) and 7th layer (blue). **(f)** A 3D model to show the spatial relation between the spin-polarized tip and the sample spins. **(g)** The upper panel is a dI/dV map across the 7 steps at -26 mV bias setting (same region as the dashed rectangle in panel (a)); the lower panel is a dI/dV spectrum linecut taken across the same region. STM setup condition: (a) $I_{set}$ = 10 pA, $V_{sample}$ = 400 mV; (c, e, g) $I_{set}$ = 800 pA, $V_{sample}$ = 200 mV, $V_{exc}$ = 2 mV.

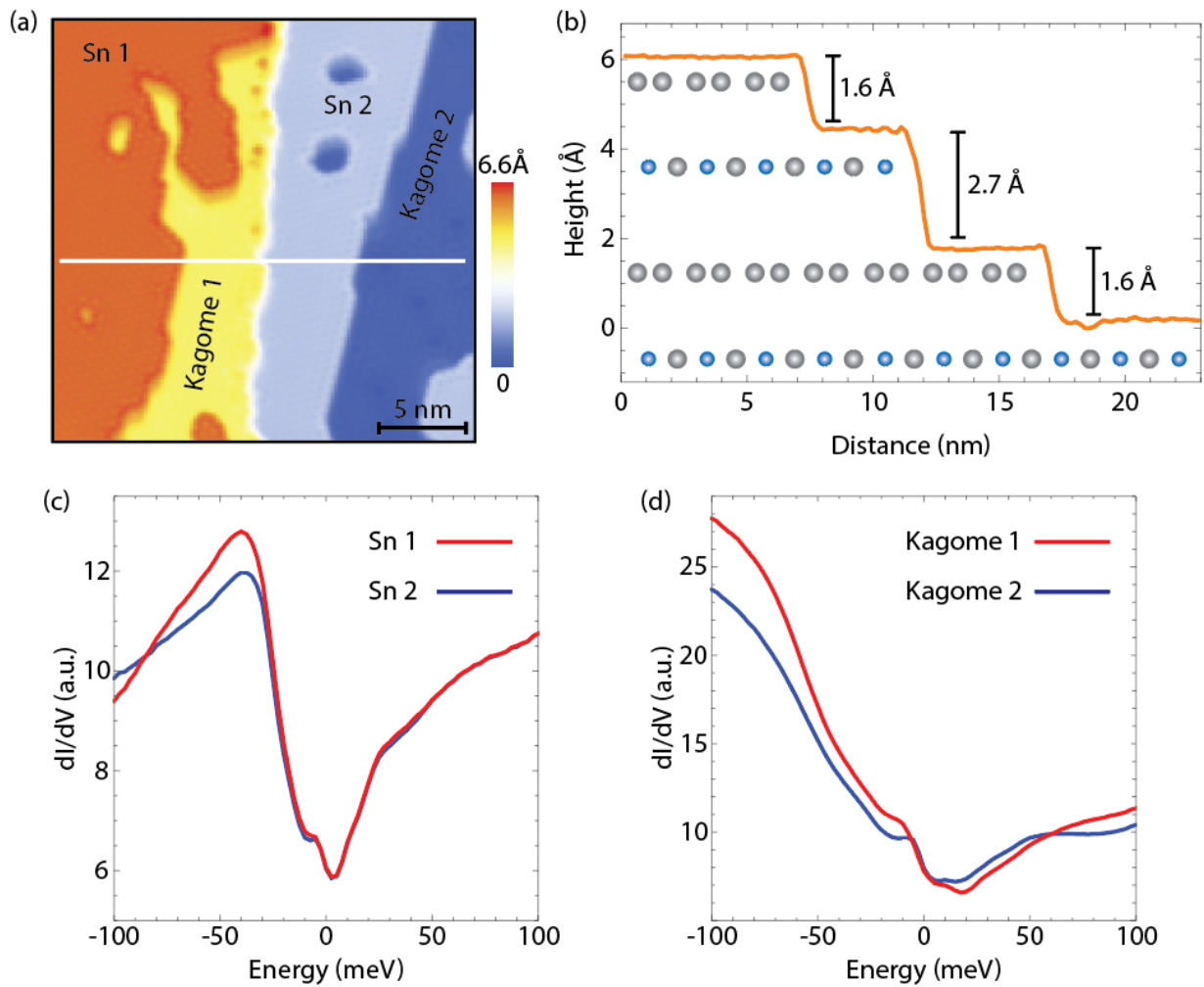

**Figure 4. Spin-polarized imaging of different surface terminations.** **(a)** STM topograph showing consecutive Sn-Fe$_3$Sn-Sn-Fe$_3$Sn terraces. **(b)** Topographic profile along the white line denoted in (a). **(c,d)** Average dI/dV spectra acquired on the inequivalent (c) Sn and (d) Fe$_3$Sn terminations. STM setup conditions: (a) I$_{set}$ = 200 pA, V$_{sample}$ = -45 mV; (c,d) I$_{set}$ = 600 pA, V$_{sample}$ = 100 mV, V$_{exc}$ = 2 mV.

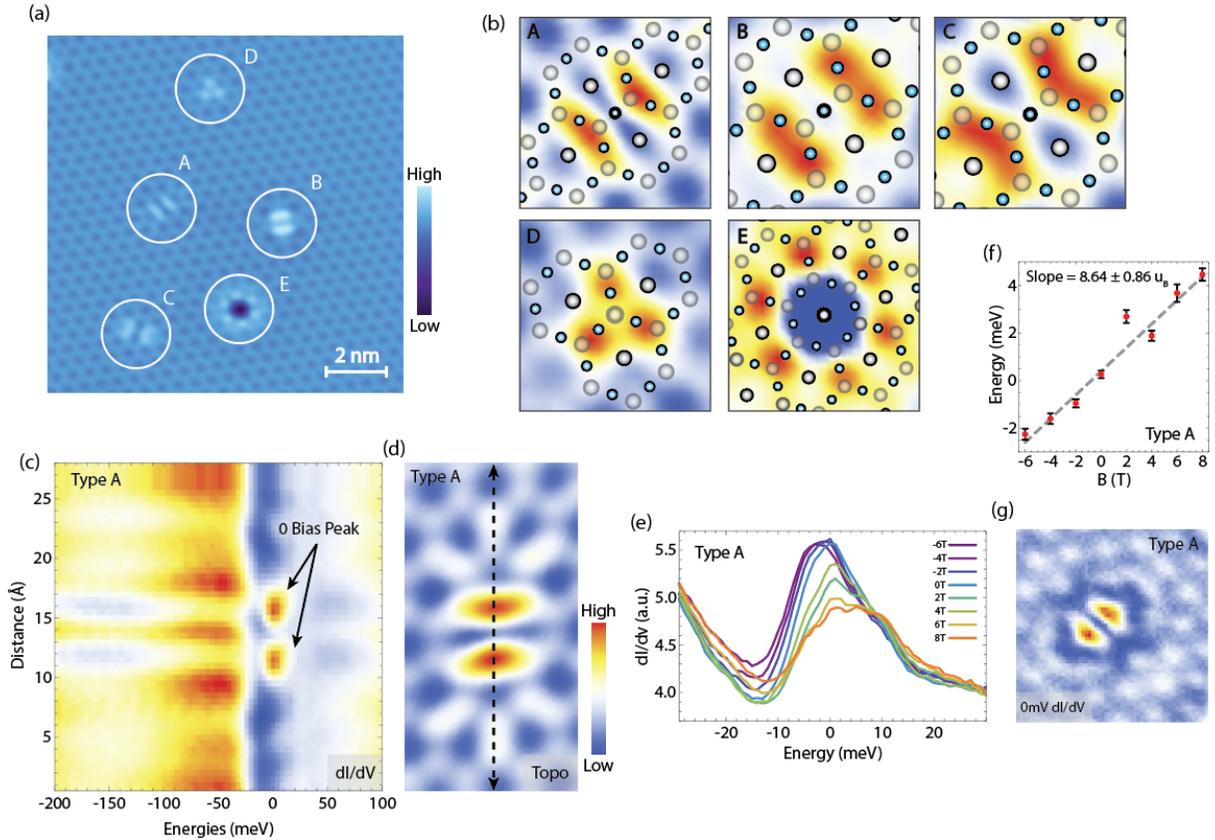

**Figure 5. Impurities and field-tunable bound states.** (a) STM topograph of the Sn surface containing several types of impurities labeled by different letters. (b) Zoom-ins on individual impurities circled in (a) with the lattice structure superimposed. (c) A linecut taken across a type A impurity center along its long axis of symmetry shown in (d). (d) A topograph of type A impurity. (e) dI/dV spectra taken over the bright resonances on either side of the center of impurity A under magnetic fields from -6 T to 8 T applied perpendicular to the sample (minus sign denotes the reversal of magnetic field direction). (f) Fitted peak positions in (e) plotted as a function of magnetic field. The slope extracted from this linear dispersion is 8.64 ± 0:86 $\mu_B$. (g) dI/dV map at 0 mV of impurity A in (a). STM setup conditions: (a,b) $I_{set}$ = 100 pA, $V_{sample}$ = 20 mV; (c-g) $I_{set}$ = 600 pA, $V_{sample}$ = 100 mV, $V_{exc}$ = 3 mV.

# References

1. Sachdev, S. Kagome- and triangular-lattice Heisenberg antiferromagnets: Ordering from quantum fluctuations and quantum-disordered ground states with unconfined bosonic spinons. *Phys. Rev. B* **45**, 12377–12396 (1992).

2. Mazin, I. I. *et al.* Theoretical prediction of a strongly correlated Dirac metal. *Nat. Commun.* **5**, 4261 (2014).

3. Guo, H.-M. & Franz, M. Topological insulator on the kagome lattice. *Phys. Rev. B* **80**, 113102 (2009).

4. Bilitewski, T. & Moessner, R. Disordered flat bands on the kagome lattice. *Phys. Rev. B* **98**, 235109 (2018).

5. Neupert, T., Santos, L., Chamon, C. & Mudry, C. Fractional Quantum Hall States at Zero Magnetic Field. *Phys. Rev. Lett.* **106**, 236804 (2011).

6. Sun, K., Gu, Z., Katsura, H. & Das Sarma, S. Nearly Flatbands with Nontrivial Topology. *Phys. Rev. Lett.* **106**, 236803 (2011).

7. Tang, E., Mei, J.-W. & Wen, X.-G. High-Temperature Fractional Quantum Hall States. *Phys. Rev. Lett.* **106**, 236802 (2011).

8. Balents, L., Fisher, M. P. A. & Girvin, S. M. Fractionalization in an easy-axis Kagome antiferromagnet. *Phys. Rev. B* **65**, 224412 (2002).

9. Yan, S., Huse, D. A. & White, S. R. Spin-Liquid Ground State of the S = 1/2 Kagome Heisenberg Antiferromagnet. *Science* **332**, 1173–1176 (2011).

10. Balents, L. Spin liquids in frustrated magnets. *Nature* **464**, 199–208 (2010).

11. Yin, J.-X. *et al.* Negative flat band magnetism in a spin–orbit-coupled correlated kagome magnet. *Nat. Phys.* **15**, 443–448 (2019).

12. Kang, M. *et al.* Topological flat bands in frustrated kagome lattice CoSn. *Nat. Commun.* **11**, 4004 (2020).

13. Yin, J.-X. *et al.* Quantum-limit Chern topological magnetism in TbMn6Sn6. *Nature* **583**, 533–536 (2020).

14. Liu, D. F. *et al.* Magnetic Weyl semimetal phase in a Kagomé crystal. *Science (80-. ).* **365**, 1282–1285 (2019).

15. Morali, N. *et al.* Fermi-arc diversity on surface terminations of the magnetic Weyl semimetal $Co_3Sn_2S_2$. *Science* **365**, 1286–1291 (2019).

16. Ortiz, B. R. *et al.* $CsV_3Sb_5$: A Z2 Topological Kagome Metal with a Superconducting Ground State. *Phys. Rev. Lett.* **125**, 247002 (2020).

17. Jiang, Y.-X. *et al.* Unconventional chiral charge order in kagome superconductor $KV_3Sb_5$.

*Nat. Mater.* **20**, 1353–1357 (2021).

18. Zhao, H. *et al.* Cascade of correlated electron states in the kagome superconductor CsV$_3$Sb$_5$. *Nature* **599**, 216–221 (2021).

19. Chen, H. *et al.* Roton pair density wave in a strong-coupling kagome superconductor. *Nature* **599**, 222–228 (2021).

20. Liang, Z. *et al.* Three-Dimensional Charge Density Wave and Surface-Dependent Vortex-Core States in a Kagome Superconductor CsV$_3$Sb$_5$. *Phys. Rev. X* **11**, 031026 (2021).

21. Li, H. *et al.* Rotation symmetry breaking in the normal state of a kagome superconductor KV$_3$Sb$_5$. *Nat. Phys.* **18**, 265–270 (2022).

22. Fenner, L. A., Dee, A. A. & Wills, A. S. Non-collinearity and spin frustration in the itinerant kagome ferromagnet Fe3Sn2. *J. Phys. Condens. Matter* **21**, 452202 (2009).

23. Wang, Q., Sun, S., Zhang, X., Pang, F. & Lei, H. Anomalous Hall effect in a ferromagnetic Fe$_3$Sn$_2$ single crystal with a geometrically frustrated Fe bilayer. *Phys. Rev. B* **94**, 075135 (2016).

24. Ye, L. *et al.* Massive Dirac fermions in a ferromagnetic kagome metal. *Nature* **555**, 638–642 (2018).

25. Yin, J.-X. X. *et al.* Giant and anisotropic many-body spin–orbit tunability in a strongly correlated kagome magnet. *Nature* **562**, 91–95 (2018).

26. Ye, L. *et al.* de Haas-van Alphen effect of correlated Dirac states in kagome metal Fe$_3$Sn$_2$. *Nat. Commun.* **10**, 4870 (2019).

27. Kang, M. *et al.* Dirac fermions and flat bands in the ideal kagome metal FeSn. *Nat. Mater.* **19**, 163–169 (2020).

28. Lin, Z. *et al.* Dirac fermions in antiferromagnetic FeSn kagome lattices with combined space inversion and time-reversal symmetry. *Phys. Rev. B* **102**, 155103 (2020).

29. Xie, Y. *et al.* Magnetic flat band in metallic kagome lattice FeSn. *ArXiv* 2103.12873 (2021).

30. Lin, Z. *et al.* Flatbands and Emergent Ferromagnetic Ordering in Fe$_3$Sn$_2$ Kagome Lattices. *Phys. Rev. Lett.* **121**, 096401 (2018).

31. Ren, Z. *et al.* Plethora of tunable Weyl fermions in kagome magnet Fe$_3$Sn$_2$ thin films. *ArXiv* 2202.04177 (2022).

32. Yamamoto, H. Mössbauer Effect Measurement of Intermetallic Compounds in Iron-Tin System : Fe$_5$Sn$_3$ and FeSn. *J. Phys. Soc. Japan* **21**, 1058–1062 (1966).

33. Giefers, H. & Nicol, M. High pressure X-ray diffraction study of all Fe–Sn intermetallic compounds and one Fe–Sn solid solution. *J. Alloys Compd.* **422**, 132–144 (2006).

34. Hartmann, O. & Wäppling, R. Muon spin precession in the hexagonal antiferromagnet

fesn. *Phys. Scr.* **35**, 499–503 (1987).

35. Shikin, A. M. *et al.* Nature of the Dirac gap modulation and surface magnetic interaction in axion antiferromagnetic topological insulator MnBi$_2$Te$_4$. *Sci. Rep.* **10**, 13226 (2020).

36. Yamaguchi, K. & Watanabe, H. Neutron Diffraction Study of FeSn. *J. Phys. Soc. Japan* **22**, 1210–1213 (1967).

37. Deng, J. *et al.* Epitaxial growth of ultraflat stanene with topological band inversion. *Nat. Mater.* **17**, 1081–1086 (2018).

38. Yin, J.-X. *et al.* Fermion–boson many-body interplay in a frustrated kagome paramagnet. *Nat. Commun.* **11**, 4003 (2020).

39. Młyńczak, E. *et al.* Fermi Surface Manipulation by External Magnetic Field Demonstrated for a Prototypical Ferromagnet. *Phys. Rev. X* **6**, 041048 (2016).

40. Han, M. *et al.* Evidence of two-dimensional flat band at the surface of antiferromagnetic kagome metal FeSn. *Nat. Commun.* **12**, 5345 (2021).

41. Enayat, M. *et al.* Real-space imaging of the atomic-scale magnetic structure of Fe$_{1+y}$Te. *Science* **345**, 653–656 (2014).

42. Hänke, T. *et al.* Reorientation of the diagonal double-stripe spin structure at Fe$_{1+y}$Te bulk and thin-film surfaces. *Nat. Commun.* **8**, 13939 (2017).

43. Manna, S. *et al.* Interfacial superconductivity in a bi-collinear antiferromagnetically ordered FeTe monolayer on a topological insulator. *Nat. Commun.* **8**, 14074 (2017).

44. Choi, S. *et al.* Switching Magnetism and Superconductivity with Spin-Polarized Current in Iron-Based Superconductor. *Phys. Rev. Lett.* **119**, 227001 (2017).

45. Zhao, H. *et al.* Atomic-scale fragmentation and collapse of antiferromagnetic order in a doped Mott insulator. *Nat. Phys.* **15**, 1267–1272 (2019).

46. Zhao, H. *et al.* Imaging antiferromagnetic domain fluctuations and the effect of atomic scale disorder in a doped spin-orbit Mott insulator. *Sci. Adv.* **7**, abi6468 (2021).

47. Hu, Y. *et al.* Real-space observation of incommensurate spin density wave and coexisting charge density wave on Cr (001) surface. *Nat. Commun.* **13**, 445 (2022).

48. Hsu, P. J., Mauerer, T., Wu, W. & Bode, M. Observation of a spin-density wave node on antiferromagnetic Cr(110) islands. *Phys. Rev. B - Condens. Matter Mater. Phys.* **87**, 1–8 (2013).

49. Heinze, S. *et al.* Real-Space Imaging of Two-Dimensional Antiferromagnetism on the Atomic Scale. *Science* **288**, 1805–1808 (2000).

50. Wiesendanger, R. Spin mapping at the nanoscale and atomic scale. *Rev. Mod. Phys.* **81**, 1495–1550 (2009).

51. Xing, Y. *et al.* Localized spin-orbit polaron in magnetic Weyl semimetal $Co_3Sn_2S_2$. *Nat. Commun.* **11**, 5613 (2020).

52. Yin, J.-X. *et al.* Spin-orbit quantum impurity in a topological magnet. *Nat. Commun.* **11**, 4415 (2020).

53. Li, H. *et al.* Manipulation of Dirac band curvature and momentum-dependent g factor in a kagome magnet. *Nat. Phys.* (2022). doi:10.1038/s41567-022-01558-3

**Competing Interests**

The Authors declare no Competing Financial or Non-Financial Interests.

**Author Contributions**

STM experiments and data analysis were performed by H.L. and H.Z. Q.W. and Q.Y. synthesized and characterized FeSn bulk single crystals under the supervision of H.C.L. Z.R. and S.S. synthesized FeSn thin films. Z.W. provided theoretical input on the interpretation of data. H.L. and I.Z. wrote the paper, with the input from all authors. I.Z. supervised the project.

**Acknowledgements**

I.Z. gratefully acknowledges the support from DOE Early Career Award DE-SC0020130 for spin-polarized STM measurements.